\documentclass[sigconf]{acmart}

\AtBeginDocument{%
  }

\usepackage{xtab}
\usepackage{enumitem}
\usepackage{multirow}
\usepackage{tabularx}
\setcopyright{none}

\makeatletter
\long\def\@setcopyrightstatement{}
\def\@copyrightpermission{}
\def\@copyrightowner{}

\def\ps@firstpagestyle{%
  \let\@mkboth\@gobbletwo
  \let\@oddhead\@empty
  \let\@evenhead\@empty
  \let\@oddfoot\@empty
  \let\@evenfoot\@empty
}

\gdef\acmConference@name{}
\gdef\acmConference@shortname{}
\gdef\acmConference@date{}
\gdef\acmConference@venue{}
\def\@acmBooktitle{}

\def\@acmYear{}
\def\@copyrightyear{}
\def\@acmISBN{}
\def\@acmDOI{}

\renewcommand{\footnotetextcopyrightpermission}[1]{}
\makeatother
\begin{document}

\title{Datasets for Navigating Sensitive Topics in \\Recommendation Systems}
% \renewcommand{\shorttitle}{Datasets for Navigating Sensitive Topics in Recommendation Systems}

\author{Amelia Kovacs}
\affiliation{%
  \institution{Cornell University}
  \city{Ithaca}
  \state{NY}
  \country{USA}
}
\email{ajk296@cornell.edu}

\author{Jerry Chee}
\affiliation{%
  \institution{Cornell University}
  \city{Ithaca}
  \state{NY}
  \country{USA}
}
\email{jc3464@cornell.edu}

\author{Kimia Kazemian}
\affiliation{%
  \institution{Cornell University}
  \city{Ithaca}
  \state{NY}
  \country{USA}
}
\email{kk983@cornell.edu}

\author{Sarah Dean}
\authornote{Corresponding author.}
\affiliation{%
  \institution{Cornell University}
  \city{Ithaca}
  \state{NY}
  \country{USA}
}
\email{sdean@cornell.edu}

%% \renewcommand{\shortauthors}{}

\begin{abstract}
  Personalized AI systems, from recommendation systems to chatbots, are a prevalent method for distributing content to users based on their learned preferences. However, there is growing concern about the adverse effects of these systems, including their potential tendency to expose users to sensitive or harmful material, negatively impacting overall well-being. To address this concern quantitatively, it is necessary to create datasets with relevant sensitivity labels for content, enabling researchers to evaluate personalized systems beyond mere engagement metrics. To this end, we introduce two novel datasets that include a taxonomy of sensitivity labels alongside user-content ratings: one that integrates MovieLens rating data with content warnings from the Does the Dog Die? community ratings website, and another that combines fan-fiction interaction data and user-generated warnings from Archive of Our Own. 

\end{abstract}




\maketitle

\section{Introduction}
Over the past decade there has been growing concern in academic, public, and regulatory spaces about negative effects of personalized recommendation.
For example, fertility related advertisements have been repeatedly shown on YouTube to infertile women whom have tried to opt out~\cite{voxtargetads}, and eating disorder content has been algorithmically recommended on TikTok~\cite{mashedtik}.
These 
recommendation systems are classically designed to maximize engagement, an objective which may lead to adverse outcomes. 
While academic work has investigated bias and fairness, there has been less work on the question of sensitive or harmful content. 
Existing academic work on harm from recommendations is largely theoretical, proving properties of simplified mathematical models, or audit based, by creating fake profiles on social media sites and tracking recommendations. 

In this paper, we address the need to better understand the interplay between preference data, personalization, and sensitive or harmful content.
Benchmark datasets are crucial to enable such analysis and promote the development of better personalization strategies. 
We propose novel datasets that augment standard preference data to account for certain notions of harm.
\emph{Sensitivity labels}---including trigger warnings or content warnings---have naturally emerged on many online media platforms.
They provide a categorization system which allows different users to avoid subsets of content they find objectionable.
These labels provide an explicit way to measure a form of user harm. We emphasize that sensitive content shouldn't be outright banned; instead, users should have the agency to avoid such content, if they choose.
For example, some users may choose to avoid animal deaths in their content recommendations, so measuring the prevalence of animal deaths in recommendations provides insights into potential user harm. 
We provide datasets which allow the research community to evaluate if personalization with preference data drives increased harm measured via content warnings.

Our contributions are: 
\emph{First,} we propose two novel datasets which augment standard user-content preferences with content warning labels. 
The first dataset combines MovieLens ratings with warnings from \url{doesthedogdie.com}; the second combines fan-fiction interaction from \url{archiveofourown.org} with warnings from the Webis Trigger Warning Corpus~\cite{wiegmann2023trigger}.
\emph{Second,} we conduct descriptive analyses on our novel datasets. 
\emph{Third,} we present a preliminary analysis on the extent standard recommendation algorithms amplify sensitive labels.
\emph{Finally,} we make our datasets\footnote{Relevant data files can be found on Hugging Face: 
\url{huggingface.co/datasets/sdeangroup/NavigatingSensitivity}}
and code\footnote{We provide code and instructions to download, clean, process, and analyze the data on our GitHub: 
\url{github.com/sdean-group/Navigating-Sensitivity}}
publicly available for other researchers to build on.

\section{Related Work}
\label{sec:related_work}

\subsection{Datasets of Sensitive or Harmful Content}

There are several types of sensitive or harmful content, and several academic datasets for studying their prevalence online. 
Content which spreads misinformation or conspiracy theories has received much attention.
\cite{faddoul2020longitudinal} study the presence of conspiracy videos on YouTube over time, and \cite{liaw2023younicon} provide the YouNICon dataset of conspiracy YouTube videos
\cite{nielsen2022mumin} provide Mumin, a dataset of fact checked misinformation on Twitter.
This type of harmful content has also recieved attention from the recommender systems community:
\cite{fernandez2021analysing} and \cite{tommasel2022recommender} link misinformation labels with recommendation data to study amplification.

Another type of harmful content is hate speech or toxic content. \cite{almerekhi2020these} present a dataset of Reddit comments and toxicity labels. \cite{mollas2020ethos} present a dataset of hate speech in Reddit and YouTube comments. \cite{wulczyn2017ex} provide a dataset of personal attacks in Wikipedia comments.
For all of these works, the labels were generated by crowd workers.

The aforementioned types of harmful content are generally viewed as universally negative,
and thus datasets are created and used with the implicit goal of limiting or preventing its proliferation.
These goals can be fraught, since the precise definition of concepts like ``hate speech'' and ``misinformation'' are often contested and may suffer from biases against marginalized groups~\cite{vidgen2020directions}.

In contrast, our focus is on user generated content warnings, which are created by the same communities who use them as a tool to curate media consumption.
As a result, this form of sensitive content is less relevant to censorship and more relevant to ensuring user agency.
\cite{wiegmann2023trigger} provide the Webis Trigger Warning Corpus, an extensive dataset and taxonomy of trigger warnings which we leverage.
Despite the relevance of sensitive content labels to real world recommendation systems like Instagram~\cite{insta} and TikTok~\cite{tiktok}, we are not aware of recommendation focused datasets that provide sensitive content labels.
We hope this work will fill that gap.

\subsection{Harm in Recommendation Systems}
There is extensive work 
characterizing various types of harmful impacts from recommender systems.  
\cite{shelby2023taxonomy} propose a taxonomy of sociotechnical harms based on an extensive literature review. 
Five major types of sociotechnical harms are categorized: representational, allocative, quality of service, interpersonal, and social systems harms.

Some harms studied act at the societal level.
\cite{ribeiro2020auditing} conduct an audit study of YouTube to study radicalization pathways.
\cite{ledwich2022radical} study filter bubbles on YouTube by analyzing the recommendations of stylized bots with content preferences and watch history.
\cite{levin2021dynamics} present several works on the dynamics of political polarization; mathematical models of polarization are formulated and theoretically analyzed.
\cite{restrepo2021social} observe how Facebook parenting communities are pushed closer to extreme communities, and then use a dynamical systems model to derive strategies for controlling this type of amplification.
\cite{whittaker2021recommender} conduct an empirical analysis of YouTube, Reddit, and Gab's recommendation systems when interacting with far-right content, showing that YouTube amplifies extreme content, while Reddit and Gab do not.
\cite{gormann2022dangers} conducts theoretical and simulation studies on the consequences of failed alignment with human values.
Other harms studied act at the individual level.
\cite{lin2016association} survey adults age 19 to 32 about social media use and depression, finding that increased social media use correlates strongly with an increased odds of depression.
\cite{smith2022recommender} use case studies to summarize common causes of algorithmic harm and their negative consequences.

Work on multi-objective recommender systems has in part been developed as a response to these observed harms.
\cite{zheng2022survey} and~\cite{jannach2022multi} provide surveys of this field.
In addition to the standard engagement maximization, these multi-objective works also optimize for diversity~\cite{vargas2011rank}, fairness~\cite{xiao2017fairness}, multi-stakeholder utility~\cite{surer2018multistakeholder}, polarization~\cite{suna2021user}, and harm proxies~\cite{singh2020building}.
This line of work requires supplemental data to measure the additional objective that is co-optimized with engagement maximization.
The datasets we provide will enable multi-objective work that balances harm as measured by unwanted exposure to sensitive content.

\subsection{Evaluating Fairness, Bias, Amplification}

Unfairness and bias are other 
consequences that may result from recommendation systems.
These topics have received much attention from the academic community; 
see surveys by \cite{chen2023bias} and \cite{ekstrand2022fairness}.
Many measures of bias and (un)fairness capture discrepancies between expressed user preferences and recommendations.
Over- or under-recommending certain content categories constitutes \emph{mis-calibrated} recommendations \cite{steck2018calibrated}.
Several works identify bias relating to item popularity~\cite{abdollahpouri2019impact,ekstrand2018all}, item genre~\cite{lin2019crank}, and creator or user demographics \cite{ekstrand2018exploring,ekstrand2018all,shakespeare2020exploring}
in domains including movies, music, books.
Another line of work investigates bias via
accessibility using counterfactual metrics.
Rather than measuring 
mis-calibrated recommendations, it is measured by what \emph{could be} recommended~\cite{akpinar2022counterfactual,dean2020recommendations,curmei2021quantifying,guo2021stereotyping}.

We hope that the datasets we provide will allow for  investigations of bias and amplification as they relate to sensitive content; we present a preliminary analysis along these lines in Section~\ref{sec:experiments}.

\section{Datasets}
\label{sec:datasets}

We present two datasets to enable study of the relationship between sensitive content and recommendation systems. The datasets each contain: 1) A \textit{sensitivity table} enumerating item identifiers and associated content warnings, and 2) An \textit{interaction table} listing the rating or presence of a "like" between each user and item. 
We define \textit{Interaction Density} as the number of interactions divided by the product of the number of users and items and
\textit{Warning Density} as the number of instances that a warning is applied to a work divided by the number of warnings times the number of works.

\subsection{MovieLens and Does the Dog Die?}
\label{sec:ml_ddd}
MovieLens is a movie recommendation service which provides data on user ratings of movies,  range from 0.5 (dislike) to 5 (like) \cite{movielens}.
To study the relationship between user preferences and sensitive content, we augment the Movielens 25M dataset~\cite{movielens}.
We leverage \url{doesthedogdie.com} (DDD), a platform of community-generated trigger warnings for movies, TV shows, and other media. Users can vote "Yes" or "No" on whether a given warning applies to a work, determining sensitivity labels. Using the DDD API\footnote{We received permission to collect this data, which contains only vote totals.}, we match the IMDb and TMBD identifiers of movies in the MovieLens 25M dataset to entries on DDD.
We create a table of movies and associated vote totals for each content warning. 
We filter the rating dataset provided by MovieLens to only contain works found on DDD (52\% of the movies) and users who interacted with at least three works in the sensitivity table (100\% of users). 
We call this dataset ML-DDD.

\begin{figure}[t]
    \centering
    \includegraphics[width=0.50\linewidth]{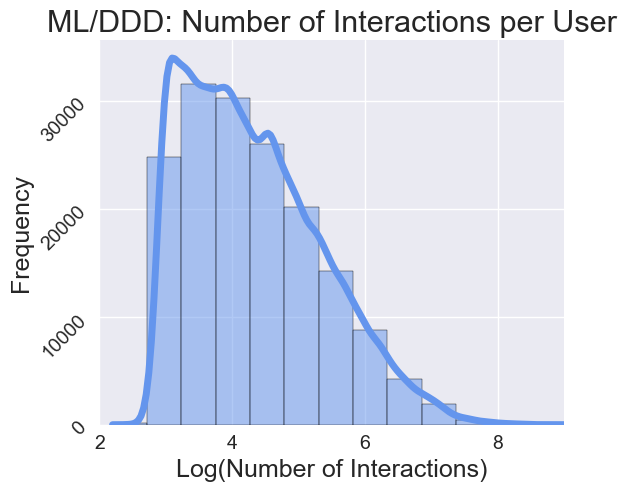}
    \includegraphics[width=0.45\linewidth]{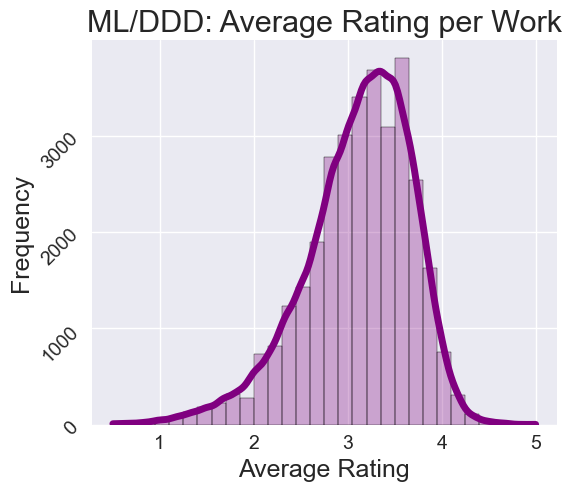}
    \includegraphics[width=0.50\linewidth]{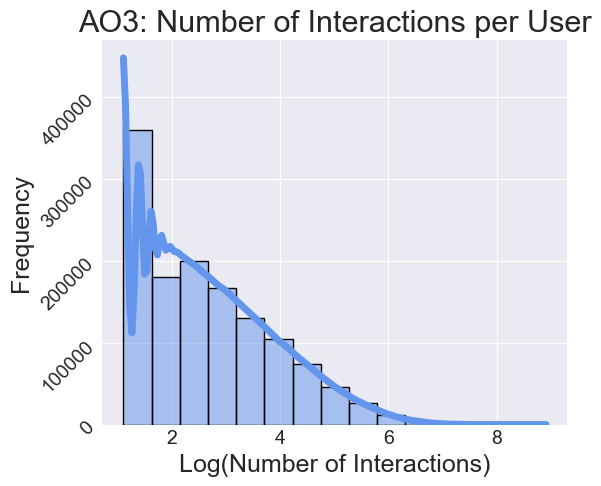}
    \includegraphics[width=0.45\linewidth]{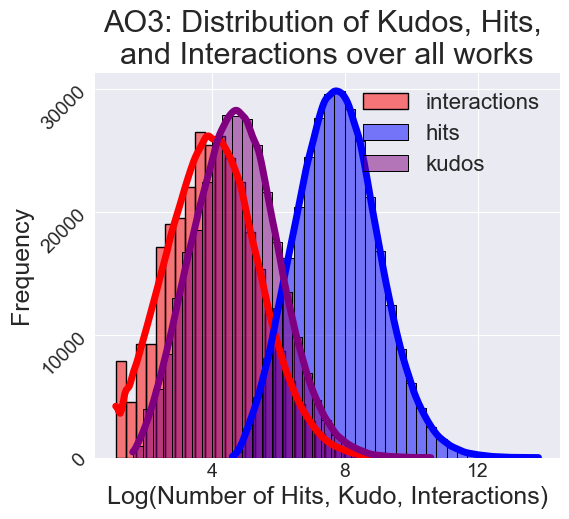}
    \caption{
    Distribution of the number of interactions per user in ML-DDD (Top Left) or AO3 (Low Left). 
    Distribution of the average rating per work for ML-DDD (Top Right), or the number of hits (reads), kudos (likes), and interactions per work in the AO3 dataset (Low Right).
    \vspace{-2em}
    }
\end{figure}

\emph{Summary Statistics.}\quad
Our dataset contains 32,604 movies, 162,541 users, and 22,867,672 interactions. On average, the movies receive 701.37 ratings and with a mean rating of 3.11.
There are 137 distinct content warnings in DDD (listed in \autoref{ddd_cats} in the appendix). 
We consider a work as having a content warning (Clear Yes) if at least 75\% of the votes are "Yes",
and without the warning (Clear No) if at least 75\% of the votes are "No".
The sensitivity table has a column for each category and warning, hot-encoded to indicate if a warning label is present. 
In total, there are 155,852 "Clear Yes" warning labels, 761,303 "Clear No", 36,080 "Unclear", and 3,513,512 "No Votes". 
We compare works with and without a warning via the "Clear Yes" and "Clear No" labels throughout our analysis. 
The dataset has an interaction density of 0.43\% and a warning density of 3.49\%.

\subsection{Archive Of Our Own}
Archive Of Our Own (AO3) is a repository of fan fiction written and read by millions of users. The works and their metadata are publicly available, such as the number of reads ("hits"), the number of likes ("kudos"), and public users who have given kudos. 
The archive has a unique
tagging system maintained by users of the site. Works are tagged with user-generated labels describing the contents of the work and often serve as a warning for potentially triggering or harmful content. \citet{wiegmann2023trigger} systematically categorize 41 million user-generated tags into 36 different trigger warning categories.
They propose the Webis Trigger Warning Corpus, a dataset of 1 million fanfiction works, metadata, and trigger warnings. 

To use this data in recommendation systems it is essential to obtain information on user interactions with works, which is lacking from the Webis Trigger Warning Corpus. We collect the publicly available usernames of those who gave kudos to each work, as well as updated kudos and hit metrics. Giving kudos to a work represents an implicit interaction.
It is worth noting that the number of identifiable users who give public kudos does not always equal the total number of kudos, as individuals are able to interact as guests. 

We collected\footnote{
We contacted AO3 support and adhered to the enforced scraping rate limits.} interactions for a subset of works from the Webis Trigger Warning Corpus (nearly 30\%) between March, 2024, and May, 2024.
Data on fandom can be considered sensitive~\cite{dym2020ethical}, so we take several steps to mitigate risks:
we collect only metadata (not content),
we exclude works with less than three publicly available user interactions and kudos by users who have interacted with less than three works in the dataset,
and we pseudonymized the usernames with unique numerical identifiers. Moving forward we will refer to this dataset as AO3.

\emph{Summary Statistics.}
Our dataset contains 306,111 works, 1,304,303 users, and 45,936,871 interactions between users and works. On average, works receive 5,468.40 hits, 265.83 kudos, and 150.07 kudos from publicly identifiable users, hereon called ``interactions".

We examine the 36 distinct ``fine closed'' warnings taxonomized in the Webis Trigger Warning Corpus (a full list is included in \autoref{ao3_cats} in the appendix). 
We consider works marked by a warning similarly to the "Clear Yes" category of ML-DDD and those without a warning as "Clear No". The AO3 sensitivity table contains a column for each warning with a 1 or 0 to indicate their presence or absence. In total there are 579,610 work-warning pairs where the warning is present in the work. The dataset has an interaction density of 0.012\% and warning density of 5.26\%.
In comparison to the ML/DDD dataset, the AO3 dataset is larger, sparser in interactions, denser in warnings, and contains implicit rather than explicit user interactions.

\begin{figure}[t]
    \centering
    \includegraphics[width=0.52\linewidth]{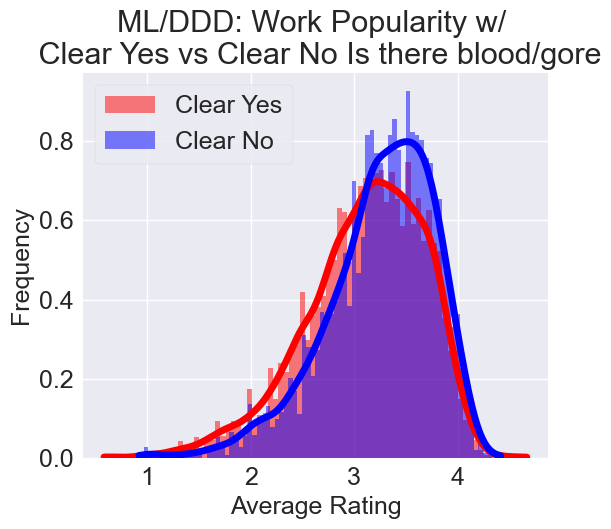}
    \includegraphics[width=0.47\linewidth]{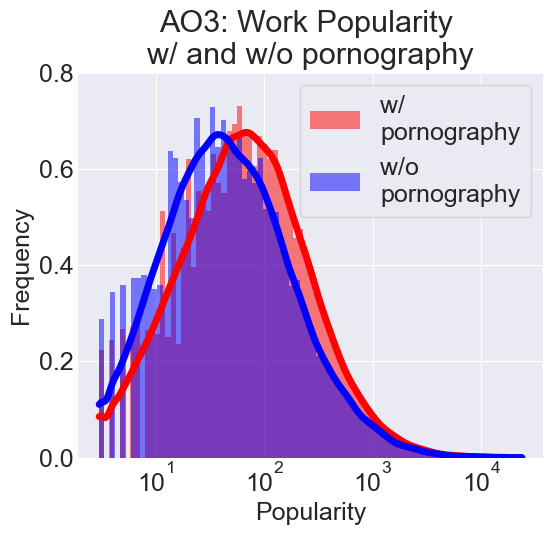}
    
    \caption{
    Distribution shift of average ratings or public kudos for works with and without a specific sensitivity label in ML-DDD (Left) or AO3 (Right).
    % (Left) Distribution shift of average ratings per work with and without the "Is there blood/gore" label in ML-DDD. (Right) Distribution shift of the number of public kudos (interactions) received by works with versus without pornography in AO3.
    \vspace{-2em}
    }
    \label{fig:ao3_pornograph_pop} \label{fig:ml-ddd_bg_pop}
\end{figure}

\section{Sensitivity Label Analysis}

We begin with a broad overview of the relationship between sensitive content and user preferences.
First, we investigate the relationship between sensitivity and popularity across various sensitive categories. 
We consider popularity in two ways: movies in ML-DDD are measured by their average rating, and works in AO3 are measured by the number of kudos received from publicly identifiable users (``interactions'' in the dataset).
Does the distribution of popularity for works marked with a content warning match the distribution of popularity for works without the warning? 
We compare the popularity distributions visually by plotting histograms of the respective metric: average ratings (ML-DDD) or interactions (AO3) in works with our without a warning label.
Figure~\ref{fig:ml-ddd_bg_pop} displays two examples for the most prevalent warnings in each dataset. 
Visually, we can see a clear shift in both the distributions.

To summarize the distribution shifts quantitatively, we perform a permutation test to examine the null hypothesis that the presence of a content warning has no correlation with work popularity. Table~\ref{fig:ml-dd_perm_test} shows the results of this test for the top 10 most prevalent content warnings in each dataset. 
Positive indicates works with the given label receive higher ratings/interactions than expected under the null hypothesis, and negative indicates less. 70\% of the ML/DDD labels and only 50\% of the AO3 have a negative correlation, meaning the collection of works with those warnings are systematically less popular.
$p$ values to determine the statistical significance of these results. 70\% of both the ML-DDD and AO3 labels have $p$ values below 0.005. We thus conclude that there is a correlation between those labels and work popularity. 

\begin{table*}[t]
\vspace{10pt}
\centering
\begin{tabularx}{.45\textwidth}{Xcc|}
\multicolumn{3}{c}{\textbf{ML-DDD}} \\
Harm & \multicolumn{2}{c}{\underline{Permutation Test}} \\
Label & Diff & P-Val \\
\midrule
blood/gore      & -0.15 & 0.00 \\
sexual          & -0.06 & 0.00 \\
\quad content   & & \\
gun violence    & -0.08 & 0.00 \\
parent death    & -0.03 & 0.01 \\
struggle        & -0.11 & 0.00 \\
\quad breath    & & \\
alcohol abuse   & 0.06  & 0.00 \\
hospital scene  & 0.03  & 0.04 \\
kidnapped       & -0.15 & 0.00 \\
sad ending      & 0.05  & 0.00 \\
drug use        & -0.02 & 0.08 \\
\end{tabularx}~\begin{tabularx}{.45\textwidth}{Xcc}
\multicolumn{3}{c}{\textbf{AO3}} \\
Harm & \multicolumn{2}{c}{\underline{Permutation Test}} \\
Label & Diff & P-Val \\
\midrule
porn       & 34.9  & 0.00 \\
violence   & 1.45  & 0.48 \\
mental     & -2.65 & 0.20 \\
\quad health & & \\
death      & -40.0 & 0.00 \\
sexual     &  19.5 & 0.00 \\
sexual     & 19.0  & 0.00 \\
\quad abuse & & \\ 
abuse      & 24.0  & 0.00 \\
medical    & 12.6  & 0.00 \\
blood/gore & -22.9 & 0.00 \\
language   & -0.28 & 0.93 \\
\end{tabularx}
\caption{
    Permutation tests results for the top 10 most prevalent ML-DDD/AO3 content warnings. 
    The permutation test examines if the presence of a content warning correlates with the average rating.
    In both datasets, 70\% of the labels have p-valueless than 0.005.
    \vspace{-2em}
    }
    \label{fig:ml-dd_perm_test}
    \label{fig:ao3_perm_test}
\end{table*}

\begin{figure}[b]
    \centering
    \includegraphics[width=0.49\linewidth]{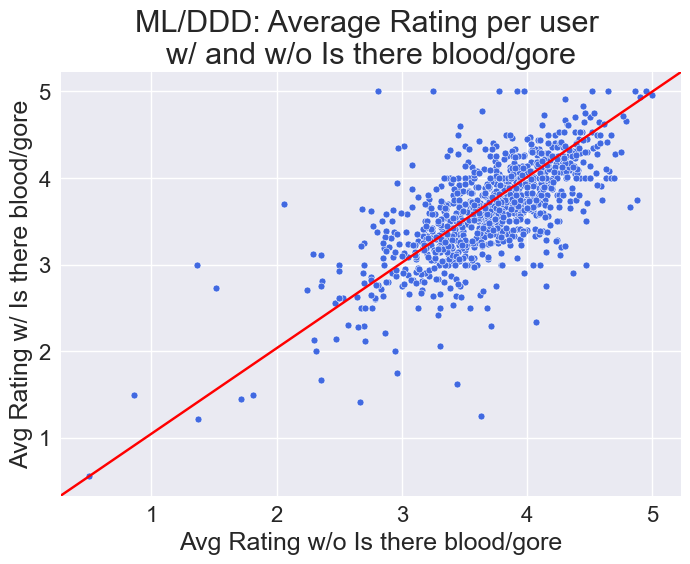}
    \includegraphics[width=0.49\linewidth]{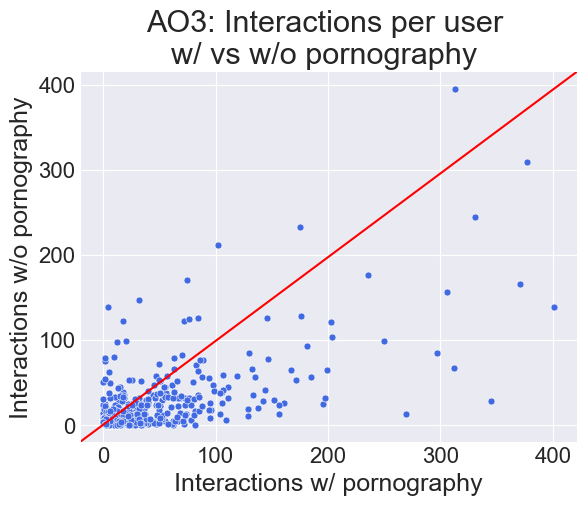}
    \caption{
    Per user average ratings or public kudos for works with and without a specific sensitivity label for ML-DDD (Left) or AO3
    (Right). 
    1,000 random users.
    More users gave higher average ratings to the sensitivity label: 57.30\% for ML-DDD and 72.60\% for AO3.
    \vspace{-2em}
    } 
    \label{fig:ao3_pornography_interactions}
    \label{fig:ml-ddd_bg_av_rating}
\end{figure}

Next, we aim to understand how individual users interact with content given the presence of warnings. We selected 1,000 random users from each dataset. Figure~\ref{fig:ml-ddd_bg_av_rating} plots the average rating (ML-DDD) or number of interactions (AO3) for each user. 
The diagonal red line has slope of one, indicating a user has equal average preference for works with and without said sensitively label. 57.30\% of the ML-DDD users fall below the red line, indicating they rate movies without the "Is there blood/gore" label higher. 
The result of the label's permutation test (Table~\ref{fig:ml-dd_perm_test}) indicates works with that content warning are rated lower on average, highlighting how individual user preferences do not always align with global popularity trends. 
Figure~\ref{fig:ao3_pornography_interactions} shows how AO3 users tend to interact more with content labeled "pornography"; 72.60\% of users fall below the red line. 

These findings are in line with the global trend (Table~\ref{fig:ao3_perm_test}). 

%%%%%%%%%%%%%%%%%%%%%%%%%%%%%%%%%%%%%%%%%%%%%%%%%%%%%%%%%%%%%%%%%%%%%%%%%%%%%%%%%%%%%%%%%%%%%%%%%%%%%%%%%%%%%%%%%%%%%%%%%%%%%%%%%%%%%%%%%%%%
\section{Amplification Analysis}
\label{sec:experiments}

The primary purpose of our novel datasets is to enable the training and evaluation of recommender systems with the notion of sensitive or harmful content in mind. We lay the foundation for this with a preliminary analysis of recommendation algorithms trained only on interaction data.

\subsection{Detailed Algorithm Descriptions}
\label{app:accuracy}

We consider two non-personalized recommendation algorithms, Random and TopPop, as baselines for each dataset \cite{Kille2019DefiningAM}. The Random algorithm generates recommendations at random, drawing works a user has not interacted with from a uniform distribution. The TopPop algorithm sorts works by popularity and recommends the top-k that a user has not seen. Popularity is measured by the average rating for a movie in ML-DDD and the number of interactions with a work in AO3.

For a more personalized approach, we also trained a Collaborative Filtering algorithm on each dataset. Collaborating filtering is a technique used in recommendation systems to predict the preferences of a user by aggregating data on other user-item interactions. It assumes that similar users will prefer similar items and uses this principle in generating ratings or recommendations. 

Singular Value Decomposition (SVD) is a well-established matrix factorization technique \cite{mnih2007svd}. It performs collaborative filtering by constructing matrices of user and item latent factors to generate predictions. We apply SVD to the ML/DDD dataset using the Python library Surprise \cite{Hug2020}. 

Unlike the explicit 0.5-5 scale ratings of ML-DDD, the AO3 dataset contains implicit user interactions from publicly available accounts awarding Kudos to a work. This means that there is no clear notion of dislike in the dataset, as it is not differentiated from works that a user simply never came across. For this reason, we use the Implicit \cite{Implicit} Python library's implementation of an Alternating Least Squares (ALS) \cite{hu2008implicit} algorithm catered for implicit data to generate recommendations for the AO3 dataset. This algorithm also uses matrix factorization for collaborative filtering, but deals with missing values by iteratively learning confidence levels to determine if the missing values indicate a negative or positive preference. 

For ML-DDD, we utilize the entire data set (22.8M interactions) to perform accuracy evaluations and generate recommendations for 1,000 random users to analyze amplification. For the evaluation of the AO3 dataset (i.e. calculating accuracy metrics), we use a representative subset of the data with 10\% of the works and their interactions (3.9M interactions) due to computational constraints. However, we used the complete dataset to train a model to generate recommendations for 1,000 random users and perform the amplification analysis.

For SVD, the algorithm's factors, epochs, learning rate, and regularization were tuned to optimize rating prediction Root Mean Squared Error (RMSE) on a 90-10 train-test split. 
For Implicit matrix factorization, the algorithm's factors, regularization, and weight of positive samples are tuned to optimize recall at k=50 recommendations. 
Hyper-parameters for both datasets can be found in the included code.

\subsection{Warning Amplification Metric}

When considering the relationship between recommendation systems and content warnings, we introduce a novel metric called \textit{Warning Amplification} inspired by metrics for algorithmic amplification \cite{ferenc2021amplification, bouchad2024amplification}.
We define
$Amplification@k$ as: for a given user, the fraction of items in their $k$ top recommendations which have a given warning divided by the fraction of items in their history with the warning.
We calculate this metric for each user in the randomly selected subset.
We normalize this metric by subtracting 1, so 0\% amplification indicates the user receives the same amount of warnings in their recommendations as exists in their history. To avoid dividing by zero in the case when the user does not have any items with the warning in their history, 
we hallucinate a single item with the given warning in the user's history.

\section{Results}
\subsection{Accuracy Evaluation}

\begin{table*}[h]
% \tiny
\small
\centering

\begin{tabular}{lrrrrrrrrr}
\toprule
 & \multicolumn{1}{l}{\begin{tabular}[c]{@{}l@{}}Precision\\ @k=10\end{tabular}} & \multicolumn{1}{l}{\begin{tabular}[c]{@{}l@{}}Precision\\ @k=50\end{tabular}} & \multicolumn{1}{l}{\begin{tabular}[c]{@{}l@{}}Precision\\ @k=100\end{tabular}} & \multicolumn{1}{l}{\begin{tabular}[c]{@{}l@{}}Recall\\ @k=10\end{tabular}} & \multicolumn{1}{l}{\begin{tabular}[c]{@{}l@{}}Recall\\ @k=50\end{tabular}} & \multicolumn{1}{l}{\begin{tabular}[c]{@{}l@{}}Recall\\ @k=100\end{tabular}} & \multicolumn{1}{l}{\begin{tabular}[c]{@{}l@{}}F1\\ @k=10\end{tabular}} & \multicolumn{1}{l}{\begin{tabular}[c]{@{}l@{}}F1\\ @k=50\end{tabular}} & \multicolumn{1}{l}{\begin{tabular}[c]{@{}l@{}}F1\\ @k=100\end{tabular}} \\
\midrule
\underline{\textbf{ML-DDD}} \\
\textbf{Random} & 1.00e-4 & 1.20e-4 & 4.00e-5 & 5.00e-4 & 2.67e-3 & 2.17e-3 & 1.67e-4 & 2.30e-4 & 7.85e-5 \\
\textbf{TopPop} & 1.00e-4 & 2.00e-5 & 2.00e-5 & 5.00e-4 &  5.00e-4 & 7.50e-4 & 1.67e-4 & 9.10e-5 & 9.40e-5 \\
\textbf{SVD} & 2.10e-3 & 4.40e-4 & 2.20e-4 & 8.48e-3 & 8.82e-3 & 8.82e-3 & 3.37e-3 & 8.38e-4 & 4.29e-4 \\
\midrule
\underline{\textbf{AO3}} \\
\textbf{Random} & 4.64e-5 & 4.96e-5 & 4.86e-5 & 3.17e-4 & 1.71e-3 & 3.29e-3 & 8.09e-5 & 9.64e-5 & 9.59e-5 \\
\textbf{TopPop} & 4.82e-4 & 7.26e-4 & 6.69e-4 & 1.43e-3 & 5.99e-3 & 1.07e-2 & 7.22e-4 & 1.29e-3 & 1.26e-3 \\
\textbf{ALS} & 4.27e-4 & 1.00e-4 & 5.14e-5 & 2.13e-3 & 2.51e-3 & 2.57e-3 & 7.11e-4 & 1.93e-4 & 1.01e-4 \\
\bottomrule
\end{tabular}
\caption{Classification Accuracy Results for each algorithm.}
\label{tab:classification_accuracy}
\end{table*}

\subsubsection{Accuracy Evaluation Metrics}

To evaluate the performance of the algorithms, we draw from the standard predictive and classification accuracy metrics~\cite{herlocker2004evaluating}. For predictive accuracy, we report the Root Mean Squared Error (RMSE) of SVD, the only algorithm which produces rating predictions. RMSE measures the square root of the average squared difference between the 0.5-5.0 scale ratings predicted by the algorithm and the true ratings for user-movie interactions. We split the rating data into a 90-10 train-test split and report the RMSE of the predictions on the test set. The split is made with respect to each user: 90\% of the user's interactions are in the trainset, and 10\% in the testset. 

However, the interaction data in AO3 is implicit: a 1 indicates a user liked an item, and a 0 indicates a user either does not like or has not seen an item. Rather than predicting this binary interaction, the algorithms trained on the AO3 data give recommendations directly. The TopPop and Random algorithms for ML-DDD give recommendations directly as well. To evaluate these algorithms, we report the classification accuracy metrics Precision, Recall, and F1 Metric at k recommendations for three values of k. We adhere by the definitions of these metrics in \cite{herlocker2004evaluating}, summarized here:
\begin{itemize}
    \item $Precision@k$. Fraction of relevant items in the k recommendations divided by k.
    \item $Recall@k$. Fraction of relevant items in the k recommendations divided by the total number of relevant items.
    \item $F1@k$. Two times the product of $Precision@k$ and $Recall@k$ divided by the sum of $Precision@k$ and $Recall@k$.
\end{itemize}

To define relevancy and calculate the classification metrics, we use a methodology similar to \cite{sarwar2000ecommerce, wilson2014improving}. We split the rating data into a 90-10 train-test split with respect to each user: again, 90\% of the user's interactions are in the trainset, and 10\% in the testset. After training each algorithm on the trainset, we generate k recommendations for each user that are unseen in their trainset interactions. The recommendations are then evaluated for precision and recall against the user's relevant testset interactions. A relevant work in AO3 is one that appears in a user's history, as all their interactions are positive. For ML/DDD, we consider movies in user's history that are rated in their top-quartile of interactions as relevant, following suit with \cite{Basu1998RecommendationAC}. 
\subsection{Main Amplification Results}

\begin{figure}[t]
    \centering
    \includegraphics[width=0.49\linewidth]{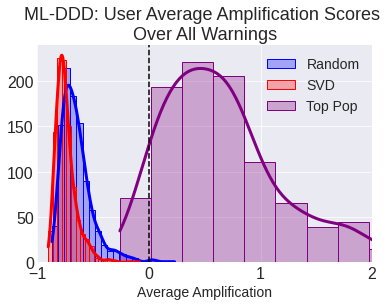}
    \includegraphics[width=0.49\linewidth]{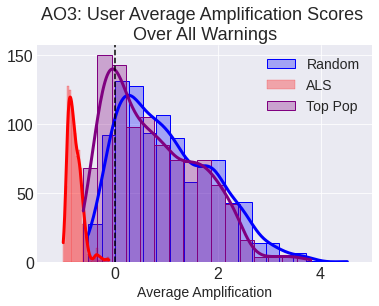}
    \caption{
    Distribution of amplification scores (ratio of warning prevalence in recommendations vs. history) averaged over all warnings in ML-DDD (Left) and AO3 (Right) for each of the 1,000 randomly selected users.
    The black dashed line is no amplification.
    In ML-DDD, TopPop produces the most amplification, and SVD the least.
    In AO3, Random produces the most amplification, and ALS the least.
    \vspace{-2em}
    } 
    \label{fig:ao3_overall_amps}
    \label{fig:ml-ddd_overall_amps}
\end{figure}

Figure~\ref{fig:ml-ddd_overall_amps} shows the distributions of average amplification scores per user for $k=100$ recommendations for each algorithm.

For ML-DDD TopPop produces the most amplification and SVD the least. 
From the permutation tests we see a high proportion of warnings correlate with popularity.
And because of the low warning density as noted in Section~\ref{sec:ml_ddd}, it is not surprising that fewer items are recommeneded with warnings.

We present the distributions of user average amplification scores for k = 100 recommendations generated by each of the algorithms in Figure~\ref{fig:ml-ddd_overall_amps}~(ML-DDD) and Figure~\ref{fig:ao3_overall_amps} (AO3). A user's average amplification is calculated by computing the mean of their Amplification@k scores for each warning. The black dashed line indicates where amplification is 0\%, i.e. there is no difference in the amount of warnings in a user's relevant history and their recommendations. In ML-DDD, we define relevant works as those in a user's top quartile of ratings. In AO3, all works in a user's history are relevant since they correspond to a positive interaction~(kudos).

In Figure~\ref{fig:ml-ddd_overall_amps} we see that TopPop produces the most amplification and SVD the least. TopPop recommends popular works to users, and as examined in the permutation tests, a high proportion of warnings correlate positively with work popularity. Given the non-personalized nature of this algorithm, users are given popular yet warning-dense recommendations. The Random recommender chooses works at random, and given the low warning density of the dataset as noted in Section 3.1, it is not surprising that the algorithm produces less recommendations with warnings than those in a user's relevant history. The SVD recommender has the lowest average amplification scores and recommends less items with a warning than in a user's relevant history. 

In Figure~\ref{fig:ao3_overall_amps} we see the user average amplifications for AO3. The Random algorithm produces the most amplification and ALS the least. The Random algorithm likely amplifies warnings more than in ML-DDD given the higher warning density of the AO3 dataset. Similarly to the ML-DDD analysis, the personalized recommender amplifies warnings the least, consistently deamplifying sensitive content.

\begin{figure}[ht]
    \centering

    \includegraphics[width=0.4\linewidth]{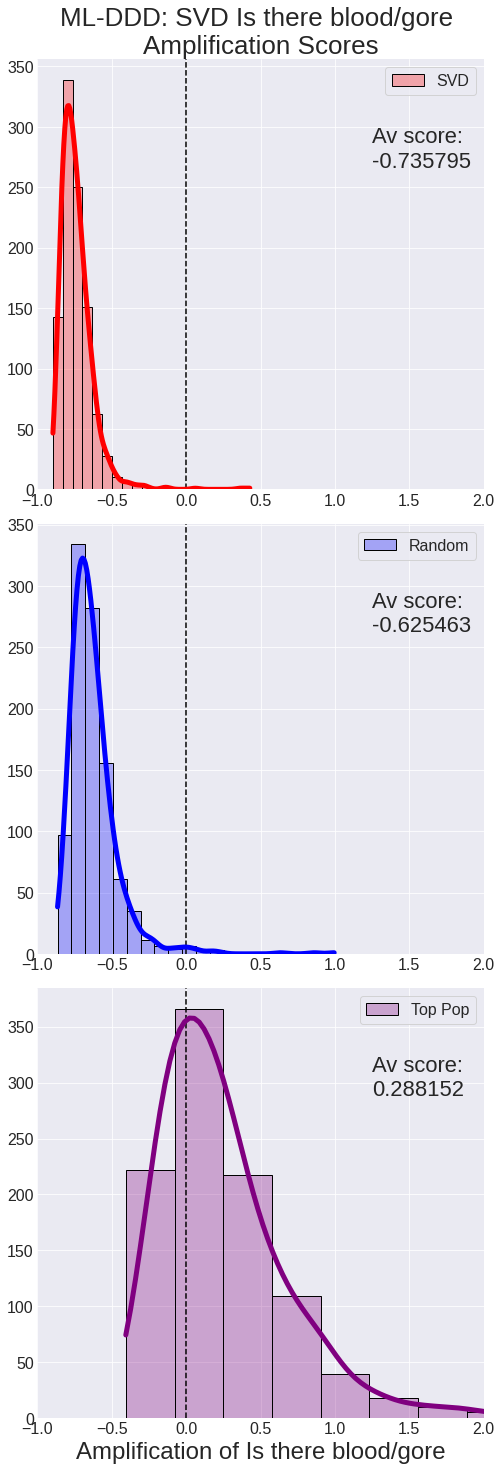}
    \includegraphics[width=0.38\linewidth]{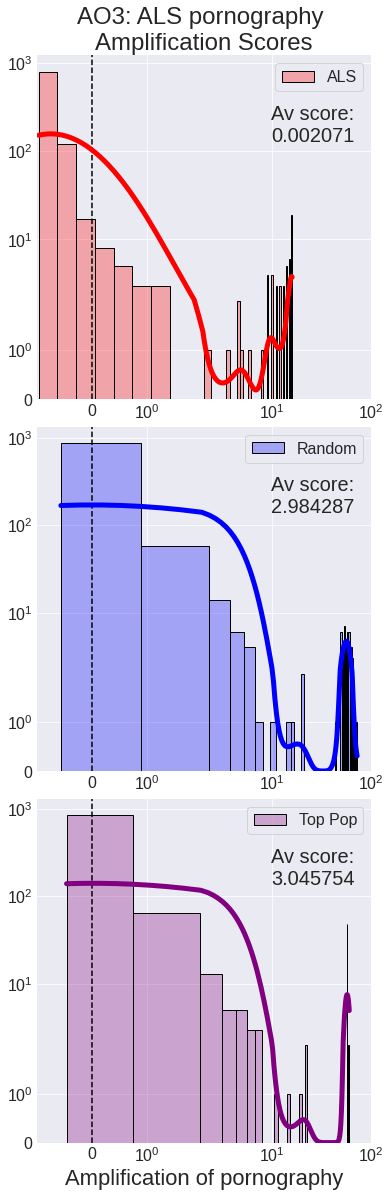}
    \caption{Amplification of the ML-DDD "Is there blood/gore" warning by the Random, TopPop, and SVD algorithms for k = 100 recommendations. The black dashed line represents 0\% amplification, indicating the percent of works with blood/gore is the same in a user's history and their recommendations.}
    \label{fig:blood_gore_amps}
\end{figure}

We also consider the average amplification of each warning, calculated by computing the mean of all user's Amplification@k scores for the single warning. In Figure~\ref{fig:blood_gore_amps} we see a comparison of the amplification of "Is there blood/gore" by the ML-DDD algorithms and "pornography" by the AO3 algorithms. Following suit with the user average amplification analysis for ML-DDD, TopPop produces the most amplification of the blood/gore warning and SVD the least. The user average amplification trends for AO3 are also closely replicated on the pornography warning, but with TopPop amplifying slightly more than Random. From the permutation tests, we know blood/gore correlates negatively with popularity and pornography correlates positively. This is reflected in the figures as amplification for blood/gore for all algorithms is much lower than that of pornography.

\subsubsection{Accuracy Results}

\textit{Predictive Accuracy}. The SVD algorithm has a RMSE of 0.7625 when trained then tested on the 90-10 random split of the data as described in Section 4.2. To generate novel recommendations for the 1,000 randomly selected users in the ML-DDD dataset, we train SVD on the entire dataset and evaluate its performance on the interactions of the random users. This produced a RMSE of 0.6733.

\textit{Classification Accuracy}. The Precision, Recall, and F1 metrics at k~=~10, 50, and 100 recommendations are displayed in ~\autoref{tab:classification_accuracy}. The Random algorithms perform significantly worse than both personalized recommenders on all metrics, which is expected of a simple baseline which does not consider any aspects of the data. In ML-DDD, the Random algorithm slightly outperforms TopPop, which may be due to the high interaction density of the dataset. Conversely, the TopPop algorithm performs quite well in AO3, outdoing both the Random and ALS algorithms. This indicates the users of AO3 are more likely to have interacted with popular content. SVD achieves higher accuracy than ALS, potentially due to the more telling explicit interactions of ML-DDD.

\subsection{Conclusion}
Overall, our preliminary analysis of the relationship between sensitive categories and recommendation systems reveals how the personalized algorithms do not amplify warnings in comparison to baselines. This finding presents an interesting avenue for future research to explore, aiming to understand the nuances in the algorithms that lead to deamplification.

\section{Discussion}
\label{sec:discussion}

In this paper, we present two novel datasets to enable understanding the interplay between preference data, personalization, and sensitive or harmful content.
The datasets include both user-content interaction data and information about sensitive content categories.

We hope this data presents opportunities for future work.
One important direction is to extend our preliminary analysis to understand how (and whether) recommendation algorithms amplify sensitive content.
Another interesting set of questions arises from sensitive label disagreement, for example by investigating disagreement in user votes in DDD.
Lastly, there are many open questions around how to design recommendation algorithms that take sensitive content seriously; for example by modelling negative preferences or providing richer user controls.
It is important to address the individual-level harms that occur when unwanted sensitive content is recommended.
We hope that these datasets will spur the research community to tackle this problem.
\vspace{-0.5em}
\section*{Acknowledgements}
This work was partly funded by NSF CCF 2312774, NSF OAC-2311521, a gift from the LinkedIn-Cornell Bowers CIS Strategic Partnership, a PCCW Affinito-Stewart Award, and a Clare Booth Luce Undergrad Research Award.

\small
\bibliographystyle{ACM-Reference-Format}
\bibliography{refs}

\appendix

\section{Additional Information}
\begin{table}[!h]
\small
\caption{The 36 warning labels used in the AO3 dataset and the number of works marked with the warning.}
\begin{tabular}{rr}
\toprule
\textbf{Warning} & \textbf{\begin{tabular}[c]{@{}r@{}}\#Works\end{tabular}} \\
\midrule
pornography & 174391 \\
violence & 42084 \\
other-mental-health & 40560 \\
death & 36123 \\
other-sexual & 35784 \\
sexual-abuse & 35005 \\
abuse & 32111 \\
other-medical & 21428 \\
blood-gore & 19299 \\
abusive-language & 18708 \\
suicide & 13113 \\
childbirth & 12132 \\
child-abuse & 11449 \\
mental-illness & 11344 \\
addiction & 11179 \\
incest & 9864 \\
homophobia & 8392 \\
self-harm & 7628 \\
kidnapping & 7561 \\
other-aggression & 5717 \\
collective-violence & 5351 \\
procedures & 4784 \\
dysmorphia & 3525 \\
other-pregnancy & 1984 \\
other-abuse & 1613 \\
sexism & 1606 \\
other-discrimination & 1595 \\
racism & 1159 \\
miscarriage & 930 \\
animal-abuse & 744 \\
transphobia & 705 \\
abortion & 561 \\
religious-discrimination & 398 \\
ableism & 390 \\
classism & 326 \\
body-shaming & 67 \\
\bottomrule
\end{tabular}
\label{ao3_cats}
\end{table}

\newpage
{\small
\topcaption{}

\tablecaption{The 137 Does the Dog Die? trigger warnings used in the ML-DDD dataset along with the number of works marked "Clear Yes" for the warning and number of works marked "Clear No".}\label{ddd_cats}

\tablefirsthead{\toprule \textbf{Warning}&\multicolumn{1}{c}{\textbf{\#Works}} \\ \midrule}
\tablehead{%
\multicolumn{2}{c}%
{{\bfseries  Continued from previous column}} \\
\toprule
Warning&\multicolumn{1}{c}{\#Works}\\ \midrule}
\tabletail{%
\midrule \multicolumn{2}{r}{{Continued on next column}} \\ \midrule}
\tablelasttail{%
\\\midrule
\multicolumn{2}{r}{{Concluded}} \\ \bottomrule}
\begin{xtabular}{ll}
Is there blood/gore & 6121 \\
Is there sexual content & 5406 \\
Is there gun violence & 4588 \\
Does a parent die & 3743 \\
Does someone struggle to breathe & 3372 \\
Does someone abuse alcohol & 3089 \\
Is there a hospital scene & 3074 \\
Is someone kidnapped & 3045 \\
Does it have a sad ending & 2980 \\
Does someone use drugs & 2935 \\
Is there a dead animal & 2822 \\
Is someone tortured & 2762 \\
Does someone fart or spit & 2756 \\
Does a car crash & 2747 \\
Is someone stalked & 2734 \\
Is there shaving/cutting & 2630 \\
Does someone cheat & 2617 \\
Is someone sexually assaulted & 2557 \\
Does an animal die & 2495 \\
Are there flashing lights or images & 2395 \\
Does someone vomit & 2373 \\
Is there hate speech & 2370 \\
Are there anxiety attacks & 2293 \\
Is there shakey cam & 2215 \\
Is a child abused & 2169 \\
Does a kid die & 2167 \\
Are there jumpscares & 2161 \\
Is there domestic violence & 2145 \\
Does someone break a bone & 2076 \\
Is someone restrained & 2027 \\
Is someone gaslighted & 2009 \\
Are needles/syringes used & 1961 \\
Is there a claustrophobic scene & 1925 \\
Is there audio gore & 1912 \\
Is there a shower scene & 1888 \\
Is there addiction & 1864 \\
Does someone die by suicide & 1840 \\
Does a car honk or tires screech & 1734 \\
Does someone fall to their death & 1701 \\
Are there bugs & 1700 \\
Are animals abused & 1599 \\
Is there ableist language or behavior & 1599 \\
Is someone burned alive & 1508 \\
Is someone hit by a car & 1506 \\
Is someone sexually objectified & 1342 \\
Does the dog die & 1298 \\
Is there amputation & 1292 \\
Is there finger/toe mutilation & 1243 \\
Are there homophobic slurs & 1238 \\
Does someone self harm & 1229 \\
Are there fat jokes & 1208 \\
Are there nude scenes & 1114 \\
Does a head get squashed & 1109 \\
Is a mentally ill person violent & 1107 \\
Is there eye mutilation & 1085 \\
Does someone drown & 966 \\
Does a baby cry & 965 \\
Are there babies or unborn children & 944 \\
Is there excessive gore & 907 \\
Is there a hanging & 884 \\
Are there ghosts & 867 \\
Is there misophonia & 834 \\
Is there obscene language/gestures & 791 \\
Does someone suffer from PTSD & 785 \\
Are any teeth damaged & 782 \\
Is someone possessed & 721 \\
Is there a mental institution scene & 712 \\
Are there snakes & 668 \\
Are there n-words & 630 \\
Does a plane crash & 620 \\
Are there spiders & 602 \\
Is someone homeless & 598 \\
Does someone becomes unconscious & 576 \\
Is there childbirth & 574 \\
Does an LGBT person die & 544 \\
Is there genital trauma/mutilation & 539 \\
Does someone have cancer & 539 \\
Are there incestuous relationships & 539 \\
Does someone say "I'll kill myself" & 524 \\
Is there body dysmorphia & 516 \\
Does someone attempt suicide & 513 \\
Does someone have a seizure & 494 \\
Is there antisemitism & 485 \\
Does someone wet/soil themselves & 467 \\
Is there a large age gap & 465 \\
Is a child's toy destroyed & 437 \\
Is there cannibalism & 436 \\
Does a pet die & 431 \\
Are there "Man in a dress" jokes & 420 \\
Does someone fall down stairs & 410 \\
Does a cat die & 392 \\
Is the fourth wall broken & 385 \\
Does someone sacrifice themselves & 379 \\
Are there clowns & 379 \\
Is someone buried alive & 376 \\
Is someone misgendered & 364 \\
Is the R word used & 360 \\
Does someone asphyxiate & 348 \\
Does the black guy die first & 348 \\
Is there copaganda & 342 \\
Is a minor sexualized & 332 \\
Does a pregnant person die & 293 \\
Is there body horror & 281 \\
Does someone have a chronic illness & 273 \\
Is there a nuclear explosion & 266 \\
Is someone crushed to death & 263 \\
Does someone have an eating disorder & 249 \\
Does someone miscarry & 243 \\
Is a minority is misrepresented & 212 \\
Is electro-therapy used & 209 \\
Is a male character ridiculed for crying & 207 \\
Is a child abandoned by a parent & 191 \\
Is someone raped onscreen & 190 \\
Are there abortions & 182 \\
Does someone overdose & 166 \\
Are there razors & 147 \\
Is there autism specific abuse & 124 \\
Is an infant abducted & 112 \\
Is someone drugged & 105 \\
Is there dog fighting & 105 \\
Is Santa (et al) spoiled & 87 \\
Are there mannequins & 83 \\
Does a dragon die & 74 \\
Does someone have a stroke & 68 \\
Is there dementia/Alzheimer's & 53 \\
Is there Achilles Tendon injury & 52 \\
Is a baby stillborn & 50 \\
Are there 9/11 depictions & 38 \\
Are there fat suits & 20 \\
Is there bisexual cheating & 17 \\
Is there D.I.D. misrepresentation & 14 \\
Does a non-human character die & 13 \\
Does the abused become the abuser & 6 \\
Is there bestiality & 4 \\
Is there body dysphoria & 4 \\
Is there aphobia & 3 \\
\end{xtabular}%
}

\onecolumn

\end{document}